\documentclass[sigconf, nonacm]{acmart}

\author{Aritra Bagchi}
\affiliation{%
  \institution{National University of Singapore}
  \country{Singapore}}
\email{bagchi.aritra@nus.edu.sg}

\author{Bin Gao}
\affiliation{%
  \institution{National University of Singapore}
  \country{Singapore}}
\email{bingao@nus.edu.sg}

\author{James Spall}
\affiliation{%
  \institution{Lumai Ltd.}
  \country{United Kingdom}}
\email{james.spall@lumai.ai}

\author{Jakub Pietrak}
\affiliation{%
  \institution{Lumai Ltd.}
  \country{United Kingdom}}
\email{jakub.pietrak@lumai.ai}

\author{Weng-Fai Wong}
\affiliation{%
  \institution{National University of Singapore}
  \country{Singapore}}
\email{wongwf@nus.edu.sg}

\author{Tulika Mitra}
\affiliation{%
  \institution{National University of Singapore}
  \country{Singapore}}
\email{tulika@comp.nus.edu.sg}

\usepackage{amsmath}
\usepackage{algorithmic}
\usepackage{textcomp}
\usepackage{enumitem}


\begin{document}

\title{Rethinking KV Caching for LLM Decode: Free-Space Optical KV Regeneration for Reduction of Memory Traffic}

\begin{abstract}
Long-context large language model (LLM) decoding repeatedly reads the historical key–value (KV) cache, making multi-head attention (MHA) memory-bandwidth-bound. Caching layer-input activations instead halves historical-cache storage and traffic, but requires regenerating KV from each cached activation. On digital accelerators such as GPUs, this regeneration takes far longer than fetching the activation from HBM. The KV regeneration overhead outweighs the savings from reduced memory traffic, shifting decode attention from a memory bottleneck to a compute bottleneck.

We present \emph{OptiDecode}, a hybrid optoelectronic
decode-attention accelerator that regenerates historical KV
using free-space optics and computes attention digitally.
Projection weights remain resident on optical tiles, exploiting
reuse across historical tokens for high-throughput optical regeneration.
Scaling to large models introduces two challenges.
First, distributing weights across tiles requires sending the
same activations to every head group, while each readout die
produces only partial KV. \emph{Activation replay} shares one
set of electrical-to-optical converters across head groups
without costly broadcast. \emph{In-situ attention} avoids
transferring partial KV from readout dies to reducers by
computing on them locally and exchanging only partial scores,
probabilities, and partial outputs.
Second, processing an activation chunk through all dependent stages takes longer than one activation-fetch interval. \emph{Chunk pipelining} overlaps successive chunks
so that OptiDecode sustains the HBM activation-fetch rate. OptiDecode demonstrates
$1.92$--$1.99\times$ decode-attention speedup over an ideal GPU
KV-read baseline at equal peak HBM bandwidth (A100--B200),
for 32K-token contexts across evaluated 30B--175B MHA models,
approaching the $2\times$ upper bound from halving historical
memory traffic.

\end{abstract}

\maketitle

\section{Introduction}
\label{sec:introduction} 
Transformer-based large language models (LLMs) generate tokens one at a time during decoding. At each step, attention uses the new token's query to attend to the keys (K) and values (V) of all previous tokens. To avoid regenerating them, conventional accelerators store historical keys and values in high-bandwidth memory (HBM) as the key--value (KV) cache~\cite{10.1145/3600006.3613165}. However, this cache must be read again at every decode step. For multi-head attention (MHA), storing KV requires $2db$ bytes per token, where $d$ is the hidden dimension and $b$ is the number of bytes per element. As the context grows, these repeated reads dominate decode-attention latency, making decode attention memory-bandwidth-bound~\cite{flashdecoding2023}.

\begin{figure}[t]
  \centering
  \includegraphics[width=\columnwidth]{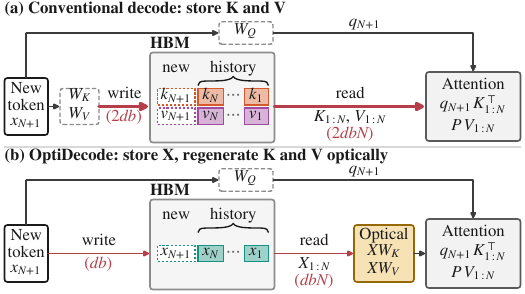}
  \caption{
  (a)~A conventional decoder caches K and V. (b)~OptiDecode caches the activation $X$
  and regenerates historical KV optically, halving HBM reads, writes, and storage.}
  \label{fig:kv-vs-x}
\end{figure}

A natural way to reduce this traffic is to \emph{cache layer-input activations instead of KV}. Both keys and values are generated from the same layer-input activations $X$, through $K=XW_K$ and $V=XW_V$, where $W_K$ and $W_V$ are key and value projection weights. Storing $X$ in an \emph{activation cache} requires only $db$ bytes per token, half the size of KV in MHA. As shown in Fig.~\ref{fig:kv-vs-x}, we can therefore halve historical-cache storage and HBM traffic by regenerating KV from the cached activations. If KV regeneration and attention sustain the HBM activation-fetch rate, decode attention can approach an ideal $2\times$ speedup.

\begin{figure}
  \centering
  \includegraphics[width=\columnwidth]{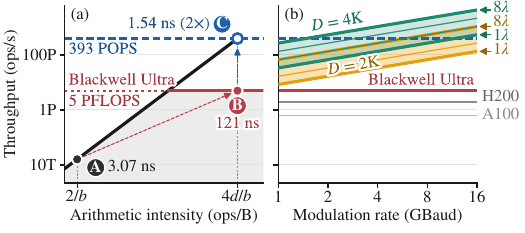}
    \caption{Trading computation for reduced memory traffic. (a) Per historical token on a single Blackwell Ultra GPU (GB300 configuration; 5\,PFLOPS dense FP8, 8\,TB/s HBM)~\cite{aubrey2025}: \textcircled{\scriptsize A} reads cached KV; \textcircled{\scriptsize B} regenerates KV from cached activations on the GPU; \textcircled{\scriptsize C} regenerates KV at the HBM activation-fetch rate. (b) The peak throughput of a single optical tile exceeds the peak dense 8-bit throughput of GPUs shown.}  
\label{fig:roofline}
\end{figure}

The difficulty is that \emph{regenerating KV requires far more computation than reading it}. Per historical token, the two projections require $4d^2$ operations (two per multiply-accumulate) to save only $db$ bytes of HBM traffic. With $d=12{,}288$~\cite{zhang2022opt}, a single NVIDIA Blackwell Ultra
GPU (GB300 configuration) fetches one 8-bit activation vector in 1.54\,ns at its 8\,TB/s peak HBM bandwidth~\cite{aubrey2025}. Regenerating its KV takes 121\,ns ($79\times$ longer), even at its 5\,PFLOPS dense FP8 peak throughput~\cite{aubrey2025}.\footnote{FP8 is this GPU's fastest 8-bit path; INT8 has the same traffic and operation count but a much lower peak. Section~\ref{sec:motivation-optical} extends the comparison to Rubin GPUs.} Caching activations on a GPU therefore replaces a memory bottleneck with a much larger computation bottleneck ($\textbf{\textcircled{\scriptsize A}}\rightarrow\textbf{\textcircled{\scriptsize B}}$ in Fig.~\ref{fig:roofline}(a)). Hiding regeneration within the 1.54\,ns activation-fetch interval (\textbf{\textcircled{\scriptsize C}},
Fig.~\ref{fig:roofline}(a)) would require about two orders of magnitude more throughput than the GPU provides. Prior work~\cite{lee2025throughput,jiang2025kvpr,zhao2024alisa,xia2025kelle,tomar2025xquant} reduces KV cache costs through selective recomputation or by overlapping recomputation with other data transfers. However, these approaches do not regenerate the entire KV history at the HBM activation-fetch rate.

\emph{Free-space optical (FSO)} computing offers a way to \emph{overcome this computation barrier.} Unlike digital accelerators, it performs matrix-vector multiplication as
light propagates in free-space: modulated light encodes the input vector, a spatial light
modulator (SLM) applies the weight matrix, and optics sum the weighted optical signals at the detector, with massive spatial parallelism~\cite{Spall:20,lumai2026iris}. A single $2{,}048\times2{,}048$ optical tile modulating input at 8\,GBaud provides 536.9\,POPS of peak throughput with eight wavelengths, far exceeding the peak dense 8-bit throughput of GPUs (Fig.~\ref{fig:roofline}(b)). More importantly, FSO computing is most efficient when many input vectors share the same weight matrix: weights stay resident on the SLM, while new input vectors stream at gigabaud rates. KV regeneration fits this pattern exactly since all historical tokens in a layer use the same projection weight matrices $W_K$ and $W_V$.

Based on this observation, we propose \textbf{OptiDecode}, a hybrid optoelectronic accelerator that caches activations, regenerates historical KV by streaming them through optical tiles holding the projection weights, and computes attention digitally (Fig.~\ref{fig:kv-vs-x}). Moving the expensive projections to optics reduces HBM traffic without making KV regeneration the new bottleneck.

However, achieving this at LLM scale raises two system challenges (elaborated in Section~\ref{sec:challenges}). 

\emph{First, the projection weights must be distributed across many optical tiles (C1).} A single optical tile cannot hold the KV projection weights of a large model. An MHA layer with hidden dimension $d=12{,}288$ has a $12{,}288\times24{,}576$~\cite{zhang2022opt} KV projection weight matrix, but an optical tile holds only $2{,}048\times2{,}048$ weights. OptiDecode therefore partitions the weights into \emph{head groups} and \emph{input slices}. This creates two problems. Each head group needs the same input activations, which are costly to broadcast (\emph{activation fan-out}, C1a). Each input slice produces partial KV on its readout die, and gathering these partial results for reduction incurs substantial die-to-die communication (\emph{partial-KV communication}, C1b).

\emph{Second, the system must sustain the HBM activation-fetch rate (C2).} Each activation chunk passes through several dependent stages (optical projection, score computation, softmax, and output computation) which together take much longer than fetching it from HBM. High optical throughput alone is therefore not enough: the whole system must accept new chunks at the rate HBM delivers them, even while earlier chunks are still in flight.

OptiDecode addresses these challenges with three techniques (Section~\ref{sec:dataflow}). \emph{Activation replay} fetches each activation chunk once and reuses a shared set of electrical-to-optical (E/O) converters to serve head groups sequentially, avoiding expensive broadcast (C1a). \emph{In-situ attention on partial KV} computes attention where partial KV are produced---directly on each readout die, and exchanges only partial scores, attention probabilities, and partial outputs, so that partial KV never leave their readout dies (C1b). \emph{Chunk pipelining} overlaps optical projection and digital attention across head groups and successive chunks, merging chunk results with online softmax~\cite{dao2022flashattention,flashdecoding2023}. With appropriately provisioned hardware  (Section~\ref{sec:architecture}), OptiDecode accepts one new activation chunk per activation-fetch interval (C2). Our major contributions are:
\begin{itemize}[leftmargin=0pt]
    \item We revisit the choice to cache KV during the decode phase, showing that
  FSO computing with resident weights makes regenerating the full KV history
  from cached activations viable at the HBM activation-fetch rate, halving
  historical-cache traffic and footprint.

    \item We identify the challenges of distributing KV projection across
    optical tiles: activation fan-out and partial-KV communication, and
    of sustaining the HBM activation-fetch rate, and design system-level techniques to address each challenge. 
    
    \item We design a heterogeneous architecture that balances optical projection, digital attention, inter-die communication, and buffering to sustain the HBM activation-fetch rate. We identify bottlenecks across models, context lengths, and HBM bandwidths, and derive the resource rates needed to sustain speedup.

    \item Our evaluation shows that OptiDecode achieves $1.92$--$1.99\times$ decode-attention speedup over the \emph{ideal GPU KV-read baseline} at equal peak HBM bandwidth (A100--B200) for 32K-token context across MHA models of 30B parameters and above. 
\end{itemize}

\section{Background}
\label{sec:background}
\subsection{Free-Space Optical Matrix-Vector Multiplication}
\begin{figure}[h]
\centering
\includegraphics[width=\columnwidth]{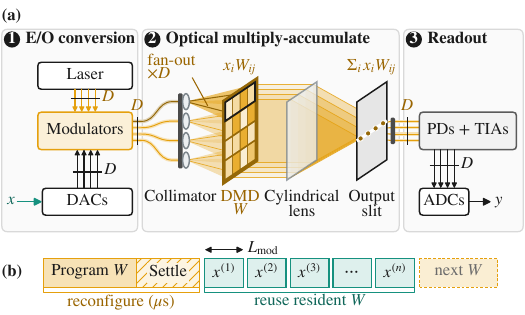}
\caption{FSO matrix-vector multiplication: (a) optoelectronic datapath;
(b) successive inputs reuse resident weights without
reconfiguring the DMD.}
\label{fig:fso-background}
\end{figure}

FSO computing uses light to perform matrix-vector
multiplication~\cite{Spall:20,lumai2026iris}. 
We write a matrix-vector multiplication as
$\mathbf{y}=\mathbf{x}\mathbf{W}$, where $\mathbf{x}$ is the
input vector, $\mathbf{W}$ is the weight matrix, and
$\mathbf{y}$ is the output. The engine comprises the following stages
(Fig.~\ref{fig:fso-background}(a)), with weights encoded on
a digital micromirror device (DMD)~\cite{TI_DLP991UUV_2025,ti2018dlp650lnir}, a type of
SLM.

\noindent\textbf{Stage 1: Electrical-to-optical conversion.} A laser is split into $D$ coherent optical channels, each driven by a digital-to-analog converter (DAC) and a modulator that encodes one signed input element $x_i$ into the optical field. The $D$ channels, therefore, form an input vector of length $D$. We call this input unit an \emph{E/O converter} and each of its channels an \emph{input channel}. A fiber array delivers these modulated signals to the optical tile.

\noindent\textbf{Stage 2: Optical multiply-accumulate.} Each input channel is expanded to illuminate one row of the DMD, allowing $x_i$ to interact with all $D$ weights in that row. The DMD modulates the incident field according to its programmed weights $\mathbf{W}$. A cylindrical lens sums the contributions along the input dimension, producing $D$ output channels with fields proportional to $y_j=\sum_{i=1}^{D}x_iW_{ij}$, all computed in parallel as light propagates. We abstract this as an \emph{optical tile} with \emph{tile dimension} $D$ holding up to $D\times D$ weights ($D^2$ elements) and computing $D$ dot products in parallel. Larger matrices must be partitioned across multiple tiles.

\noindent\textbf{Stage 3: Electronic readout.} Each output channel feeds a photodetector (PD), transimpedance amplifier (TIA), and analog-to-digital converter (ADC), which detect, amplify, and digitize the optical result. Since the computation involves signed values, the receiver must recover both sign and magnitude~\cite{Spall:20}. The ADC outputs are available for subsequent digital computation.

\subsection{Weight Residency and Optical Throughput}

FSO computing has asymmetric costs for changing weights and changing inputs. Loading a new weight matrix is \emph{slow}: it requires programming the DMD and waiting for its micromirrors to settle. Once the weights are settled, streaming new input vectors through modulators and producing outputs are \emph{fast} (Fig.~\ref{fig:fso-background}(b)).

\noindent\textbf{Input modulation rate.} Let $L_{\mathrm{mod}}$ be the interval between successive input vectors and $f_{\mathrm{mod}}=1/L_{\mathrm{mod}}$ the input modulation rate, which can reach gigabaud rates. Each interval delivers all $D$ input elements of the vector in parallel. With resident weights, $n$ input vectors therefore incur a latency of $nL_{\mathrm{mod}}$, provided the input modulation and electronic readout can sustain this rate. 

\noindent\textbf{Wavelength-division multiplexing.} With wavelength-division multiplexing (WDM)~\cite{andregg2019wavelength,feldmann2021parallel}, $N_\lambda$ wavelengths each carry a different input vector through their own DACs and modulators while sharing the same $D$ fibers and DMD pattern. Different wavelengths do not interfere coherently, so each produces its own matrix-vector result, which a demultiplexer routes to corresponding detectors.

\noindent\textbf{Optical throughput.} A $D\times D$ tile performs $D^2$ multiply accumulates per input vector, and processes $N_\lambda$ vectors per modulation interval, yielding a peak optical throughput of
\begin{equation}
P_{\mathrm{peak}}=2N_\lambda D^2 f_{\mathrm{mod}}.
\label{eq:ppeak}
\end{equation}
For example, a $2{,}048\times2{,}048$ tile at 8\,GBaud provides 67.1\,POPS with one wavelength and 536.9\,POPS with eight wavelengths (Fig.~\ref{fig:roofline}(b)). This throughput can be realized only when many inputs share one weight matrix (and avoid slow DMD reconfiguration), which is exactly the case for KV regeneration.

\subsection{Optical Fan-Out and Physical Constraints}
Optical power and precision constrain FSO computing. Within a $D\times D$ tile, each input channel must illuminate $D$ weight elements (a row of $\mathbf{W}$, Fig.~\ref{fig:fso-background}(a)). So, a larger $D$ spreads the available optical power across more elements, and a higher modulation rate leaves fewer photons per symbol; both make the target output precision harder to maintain against optical and electronic noise. More laser power compensates for this at an energy cost. WDM raises throughput without shrinking the per-wavelength photon budget, but needs more lasers, modulators, and readout channels. The choice of $D$ therefore depends on the target precision, modulation rate, and available optical power within the system's energy budget.

These physical constraints also underlie the activation fan-out
challenge introduced in Section~\ref{sec:introduction} (C1a). Even within a single optical tile, each input channel undergoes
$D$-way optical fan-out to illuminate one row of weights (Fig.~\ref{fig:fso-background}(a)).
Broadcasting the same activations across head groups adds
further fan-out across tiles, reducing the power per tile and
requiring more laser power to maintain precision.

\section{Motivation and Challenges}
\label{sec:motivation}

\subsection{Can Optical Projection Keep Pace with HBM?}
\label{sec:motivation-optical}
\begin{figure}[h]
  \centering
  \includegraphics[width=\columnwidth]{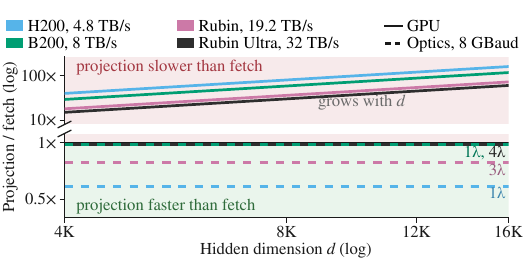}
  \caption{KV projection time relative to activation-fetch interval at equal HBM bandwidth. Even at peak throughput, GPUs (solid)
cannot keep pace with HBM fetch; optics (dashed) can.}
  \label{fig:motivation-optical}
\end{figure}

Activation caching approaches the $2\times$ upper bound on decode-attention speedup only if KV regeneration sustains the HBM activation-fetch rate. High peak optical throughput alone does not guarantee this: activation
replay serves one head group at a time (Section~\ref{sec:introduction}), so the sustained
rate is lower than the peak. Fig.~\ref{fig:motivation-optical} plots the ratio of KV projection time
to activation-fetch interval, for GPUs and FSO computing at equal peak HBM bandwidth.

GPU KV regeneration multiplies every fetched activation byte by both the key
and value projection weights, so its arithmetic intensity
grows linearly with the hidden dimension: about $49$k operations per byte for $d=12{,}288$. A
GPU's compute-to-bandwidth ratio (peak dense 8-bit throughput over peak HBM
bandwidth) is 412--911 operations per byte~\cite{nvidiaH200,nvidiaHGX,nvidiaVR200}. GPUs therefore take one to two orders of magnitude longer
to project KV than to fetch the activations, with this ratio
growing linearly with hidden dimension $d$
(Fig.~\ref{fig:motivation-optical}, solid lines). Even the Rubin Ultra's FP8 peak
requires 45 times the activation-fetch interval
for KV regeneration~\cite{nvidiaGTC2025}.

Associativity property of matrix operations could avoid per-token explicit KV regeneration by folding $W_K$ into the
query and $W_V$ into the weighted sum of activations, applied once per decode step. However, post-projection transformations such as RoPE~\cite{su2024roformer}  prevent this rearrangement for keys. Values can still be folded, but the remaining key projection alone costs about 44$\times$ the activation-fetch interval for a large model on a B200 GPU.

Optical KV projection, in contrast, can sustain the HBM activation-fetch rate. An optical tile applies its entire resident matrix, up to its tile dimension, to a token in a single modulation interval (Section~\ref{sec:background}), so even when head groups are served one at a time, projecting a token's KV finishes within the time HBM takes to fetch the activation. The required wavelength count ($N_{\lambda}$) is set by HBM
bandwidth, tile dimension, and activation precision
(Eq.~\eqref{eq:replay-freq}) and is independent of hidden
dimension. With $8$-bit activations, $2{,}048\times2{,}048$ tiles, and 8\,GBaud modulation, one wavelength suffices at H200's and B200's HBM3e bandwidth of 4.8\,TB/s~\cite{nvidiaH200} and 8\,TB/s~\cite{nvidiaHGXB200Memory}, respectively, while three and four
suffice at Rubin's HBM4 bandwidth of 19.2\,TB/s~\cite{nvidiaVR200} and Rubin Ultra's HBM4e bandwidth of 32\,TB/s~\cite{nvidiaGTC2025}, respectively (Fig.~\ref{fig:motivation-optical}).

\subsection{Challenges}
\label{sec:challenges}

\begin{figure}[t]
\centering
\includegraphics[width=\columnwidth]{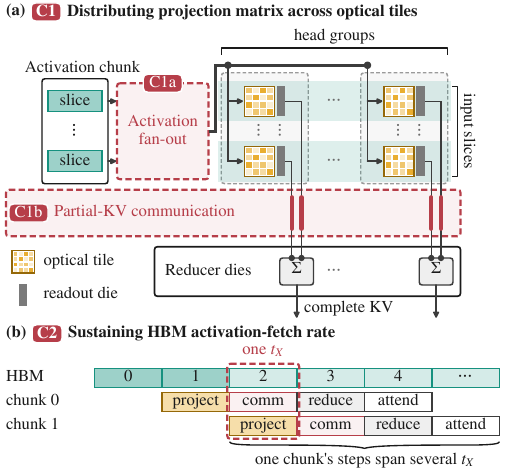}
\caption{System challenges: (a) distributed KV projection (C1)
introduces activation fan-out (C1a) and partial-KV communication
(C1b); (b) A chunk's dependent-stage latency exceeds the activation-fetch
interval (C2).}
\label{fig:challenges}
\end{figure}

Section~\ref{sec:motivation-optical} showed that optical projection can sustain the HBM activation-fetch rate. However, fast optical projection alone does not guarantee an end-to-end decode-attention speedup. A complete system must address two challenges, illustrated in Fig.~\ref{fig:challenges}.

\noindent\textbf{\textit{Challenge 1: Distributing KV projection across optical tiles (C1).}} A layer's KV projection matrix is larger than one optical tile, so we partition the matrix along both dimensions (Fig.~\ref{fig:challenges}(a)): its columns into \emph{head groups}, each responsible for the KV of a subset of attention heads, and its rows into \emph{input slices}, each covering part of the input activation vector. These two partitions introduce different data-movement costs.

\emph{Activation fan-out (C1a).} Every head group needs the same input activations, because each head's KV depends on the entire input activation vector. Broadcasting it to all head groups simultaneously is expensive: electrical broadcast requires a separate set of E/O converters per group, while optical broadcast shares the converters but splits the light across groups, making precision harder to maintain.

\emph{Partial-KV communication (C1b).} Each tile processes one input slice and produces partial KV that must be summed across slices to obtain a head group's complete KV. These partials reside on separate readout dies, and each is as large as the complete KV for its group, so gathering them at a reducer die would require substantial die-to-die communication, potentially offsetting the HBM traffic savings.

\noindent\textbf{\textit{Challenge 2: Sustaining the HBM activation-fetch rate (C2).}} To approach the $2\times$ speedup, the system must process activations at the rate at which HBM delivers them. We fetch cached activations in chunks of $C$ tokens, each chunk taking one \emph{activation-fetch interval} $t_X$ to fetch from HBM (Fig.~\ref{fig:challenges}(b)). Although optical projection can finish within $t_X$ (Section~\ref{sec:motivation-optical}), each chunk also passes through dependent stages, including communication, reduction, and attention, that together take longer than $t_X$. Processing chunks one after another would leave the HBM idle. Successive chunks must therefore overlap so that every resource completes its work for one chunk within $t_X$, even though a single chunk takes several $t_X$ intervals to finish.

\section{OptiDecode Design}
\label{sec:dataflow}
We first map the KV projection weights onto a grid of optical tiles (Section~\ref{sec:mapping}). We then address the resulting costs with \emph{activation replay} (C1a, Section~\ref{sec:activation_replay}) and \emph{in-situ attention} (C1b, Section~\ref{sec:local-attention}), and sustain the HBM activation-fetch rate with \emph{chunk pipelining} (C2, Section~\ref{sec:pipeline}).

\subsection{Projection Mapping onto Optical Tiles (C1)}
\label{sec:mapping}
\begin{figure}[h]
 \centering
 \includegraphics[width=\columnwidth]{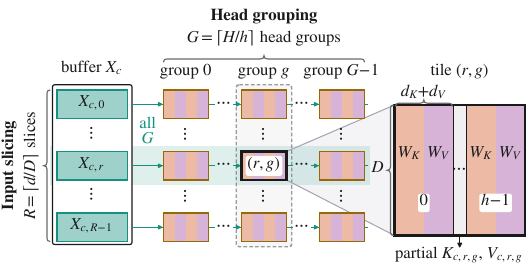}
 \caption{Projection mapping onto an $R\times G$ grid of optical tiles. Rows correspond to input slices; columns correspond
to head groups. Tile $(r,g)$ produces partial KV for
$h$ heads.}
 \label{fig:mapping}
\end{figure}

Each of the $H$ attention heads projects a token's $d$ activations into a $d_K$-element key and a $d_V$-element value. 
All heads' key and value projection weights form a projection weight matrix $W$ with one row per activation element and
$d_K+d_V$ columns per head. Because $W$ is larger than the tile dimension, we distribute $W$ across many optical tiles
(Fig.~\ref{fig:mapping}).

\noindent
\textbf{Head groups.}
We first split the columns of $W$ by attention head, keeping each head's $d_K+d_V$ columns together.
An optical tile with $D$ columns holds $h=\lfloor D/(d_K+d_V)\rfloor$ heads, which we call a \emph{head group}, so the $H$ heads form $G=\lceil H/h\rceil$ head groups (the last may hold fewer heads).
A head is never split across tiles, so it must fit in one tile ($d_K+d_V\le D$); typical heads need 128--512 columns, well within tile dimension $D$. Each column of tiles in Fig.~\ref{fig:mapping} is one head group.

\noindent
\textbf{Input slices.}
We then split the rows of $W$ into blocks of $D$ consecutive activation elements, which we call \emph{input slices}.
An optical tile with $D$ rows holds one input slice, so covering all $d$ activations ($d>D$) takes $R=\lceil d/D\rceil$ input slices (the last one is zero-padded).
Every token's activations are split at the same boundaries: $X_{c,r}$ holds the $r$-th slice of every token in chunk $c$.
Each head group therefore spans $R$ tiles, one per input slice, and each row $r$ of tiles in Fig.~\ref{fig:mapping} receives slice $X_{c,r}$ from the activation buffer.

\subsection{Activation Replay across Head Groups (C1a)}
\label{sec:activation_replay}

Every head group must project the same activation chunk, with each of its $R$
tiles receiving one input slice.  Delivering each chunk to all $G$ head groups
is the activation fan-out problem (C1a).

\begin{figure}[h]
 \centering
 \includegraphics[width=\columnwidth]{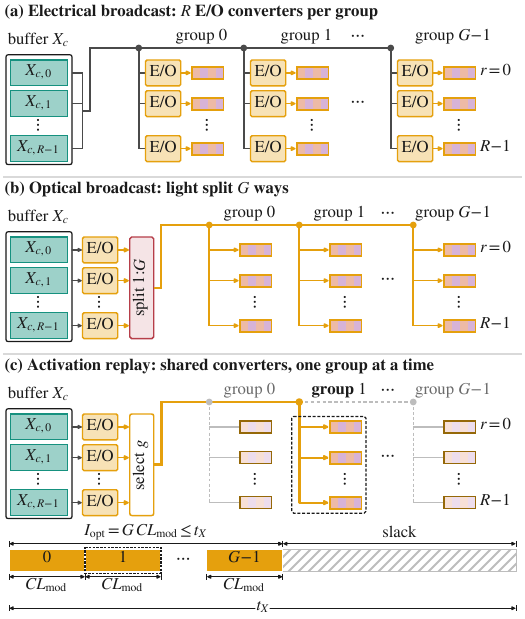}
 \caption{Delivering an activation chunk to $G$ head groups. Broadcast needs (a) $G$ times as many E/O converters, or (b) splits their light $G$ ways. (c) Activation replay shares $R$ converters across groups, one $CL_{\mathrm{mod}}$ step per group.}
\label{fig:replay}
\end{figure}

\noindent \textbf{Broadcast overheads.} Na\"ively broadcasting the chunk to all head groups at once is expensive.
Electrical broadcast gives every head group its own $R$ converters, $RG$ in
total (Fig.~\ref{fig:replay}(a)), multiplying the lasers, DACs, and
modulators, and their power by $G$. Optical broadcast keeps $R$
converters but splits their light $G$ ways (Fig.~\ref{fig:replay}(b)) on top of the fan-out within each tile (Section~\ref{sec:background}), so maintaining target precision at the detector requires about $G$ times the laser power, plus additional optical hardware to split and route the light. 

\noindent \textbf{Project one head group at a time.} OptiDecode instead serves head groups one at a
time, using time the optics would otherwise leave idle: projecting KV for a chunk's $C$ tokens takes only $C$ modulation intervals ($CL_{\mathrm{mod}}$), typically far shorter than the activation-fetch interval $t_X$. OptiDecode fetches each chunk from HBM
once into an on-chip activation buffer with $R$ banks (one per input slice), each driving one E/O
converter (Fig.~\ref{fig:replay}(c)). An optical group selector routes the $R$
modulated slices to one head group, whose $R$ tiles project them in parallel. The group selector then moves to the next group, and the activation-buffer banks are read again. The projection weights stay resident on every tile
and only the selection changes.

\noindent \textbf{Replay rate.} Replay serializes the head groups, so a chunk takes $G$
steps of $CL_{\mathrm{mod}}$ each (timeline in Fig.~\ref{fig:replay}(c)). With
$N_\lambda$ wavelengths, each step projects $N_\lambda$ tokens at once, so the
optical projection time of one chunk is
$I_{\mathrm{opt}}=GCL_{\mathrm{mod}}/N_\lambda$. Replay sustains the HBM activation-fetch rate as
long as this fits within the activation-fetch interval, i.e.,
$I_{\mathrm{opt}}\le t_X=Cdb/B_{\mathrm{HBM}}$. For MHA, the $H$ heads together have $2d$ projection columns, so
$G\approx 2d/D$, and
\begin{equation}
  I_{\mathrm{opt}} \le t_X
  \iff
  N_\lambda f_{\mathrm{mod}} \ge \frac{2B_{\mathrm{HBM}}}{bD}.
  \label{eq:replay-freq}
\end{equation}
Both the chunk size and the hidden dimension cancel: the required modulation rate
depends only on HBM bandwidth, precision, and tile dimension, which is why the
optical curves in Fig.~\ref{fig:motivation-optical}, Section~\ref{sec:motivation-optical} do not depend on $d$.

\subsection{In-Situ Attention on Partial KV (C1b)}
\label{sec:local-attention}
\begin{figure}[t]
\centering
\includegraphics[width=\columnwidth]{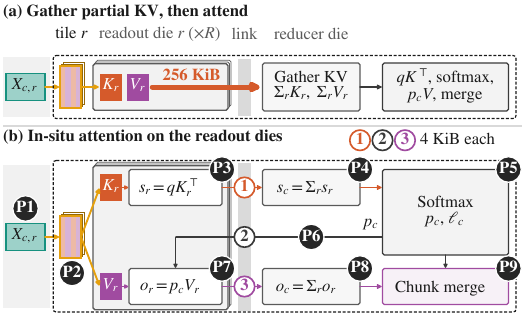}
\caption{In-situ attention for one head group (OPT-175B, per readout die per chunk). (a) The na\"ive approach gathers partial KV, requiring 256\,KiB per link. (b) Each readout die computes on its own partial KV and exchanges only partial scores \textcircled{1}, attention probabilities \textcircled{2}, and partial outputs \textcircled{3}, 4\,KiB each.}
\label{fig:local-attention}
\end{figure}

Input slicing leaves each head group's keys and values as partial KV across its $R$ readout dies. A
na\"ive approach gathers these partials over the die-to-die links, reduces them
into complete KV, and then computes attention
(Fig.~\ref{fig:local-attention}(a)). Since slicing partitions
the input dimension, not the output dimension, each readout die's
partial KV is as large as the head group's complete KV, making gathering costly. For OPT-175B with 128-token chunks, this is 256\,KiB per readout die per chunk. Moving this 256\,KiB within the 196.61\,ns needed to fetch the
next activation chunk at 8\,TB/s requires approximately
1.33\,TB/s on each link (C1b).

\noindent\textbf{Computing on partial KV.} OptiDecode instead computes attention where partial KV are produced so that partial KV never leave their readout dies. Attention uses keys and values only through the query--key (QK) and probability--value (PV) products. Both are linear, so we can apply them to partial KV and sum the results, instead of summing partial KV first. Digital engines next to the ADCs on each readout die (Section~\ref{sec:architecture}) compute on the die's own partial KV and exchange only small intermediate results with a reducer die (Fig.~\ref{fig:local-attention}(b)).

For each head and activation chunk $c$, let $K_r$ and $V_r$
denote the partial keys and values on readout die $r$, and $q$
the current query. Let $s_r$ and $o_r$ denote partial scores
and outputs; $s_c$, $p_c$, $o_c$, and $\ell_c$ denote the chunk's
complete scores, attention probabilities, output, and log-sum-exp
(LSE).

\noindent\textbf{Scores from partial keys.} Each readout die computes partial scores $s_r=qK_r^\top$ (\textbf{P3}) and sends them to the reducer die (transfer~\textcircled{1} in Fig.~\ref{fig:local-attention}(b)), which sums them (score reduction, \textbf{P4}):
\[
s_c=\sum_r s_r=\sum_r qK_r^\top=q\Bigl(\sum_r K_r\Bigr)^\top.
\]
Algebraically, this gives the same scores as first summing the digitized partial keys, yet each readout die sends one partial score per token and head instead of a partial key vector. The query is sent once per decode step and reused across chunks.

\noindent\textbf{Softmax on complete scores.} Softmax must use the complete scores $s_c$; normalizing partial scores independently would change the result. The softmax engine applies the $1/\sqrt{d_K}$ scaling and any attention mask, computes attention probabilities $p_c$ (\textbf{P5}), and retains the chunk's LSE $\ell_c$ for the later chunk merge~\cite{milakov2018online}.

\noindent\textbf{Outputs from partial values.} The attention probabilities return to every readout die (\textbf{P6}, transfer~\textcircled{2}), which computes a partial output $o_r=p_cV_r$ (\textbf{P7}) and sends it to the reducer die (transfer~\textcircled{3}) for summation (output reduction, \textbf{P8}):
\[
o_c=\sum_r o_r=\sum_r p_cV_r=p_c\Bigl(\sum_r V_r\Bigr).
\]
Algebraically, this produces the same output as first summing the digitized partial values, yet each readout die sends one partial output vector per head instead of a partial value vector per token. Partial KV never leave their readout dies; only partial scores, attention probabilities, and partial outputs are transferred between dies.

\noindent\textbf{Communication reduction.} For OPT-175B with $C=128$, each of the three messages is
4\,KiB per readout die per chunk, using INT32 scores and FP32
probabilities and outputs. Each link transfers 8\,KiB
readout-to-reducer (scores and outputs) and 4\,KiB
reducer-to-readout (probabilities). At 8\,TB/s HBM bandwidth,
matching B200, the activation-fetch interval is 196.61\,ns,
requiring only 41.7\,GB/s readout-to-reducer and 20.8\,GB/s
reducer-to-readout, compared with 1.33\,TB/s for gathering
partial KV.

\noindent\textbf{Merging attention across chunks.} Each chunk’s attention probabilities are normalized over its own tokens; OptiDecode combines chunk outputs using their LSEs~\cite{dao2022flashattention,flashdecoding2023}. For each head, the chunk-merge engine maintains a running output $o$ and LSE $\ell$. When chunk $c$ completes, it updates $\ell'=\log(e^{\ell}+e^{\ell_c})$ and $o'=e^{\ell-\ell'}o+e^{\ell_c-\ell'}o_c$ (\textbf{P9}). This online-softmax merge recovers full-history attention without reconstructing KV or retaining all attention probabilities.

\subsection{Chunk Pipelining at the Activation-Fetch Rate (C2)}
\label{sec:pipeline}
\begin{figure}[t]
\centering
\includegraphics[width=\columnwidth]{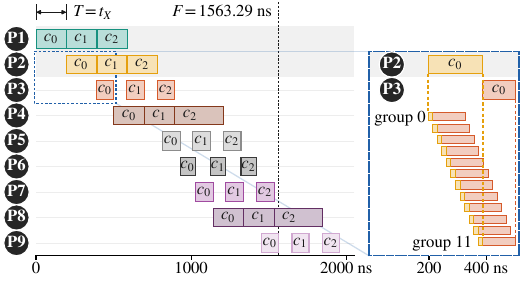}
\caption{Chunk pipelining (OPT-175B, $C=128$, 8\,GBaud,
8\,TB/s). Left: successive chunks start every $T=t_X$;
$c_0$ finishes after $F$. P2 spans all head groups;
P3--P9 show the last group. Right: zoom into the dashed box,
showing each head group starting partial score computation
(P3) as soon as its replay step (P2) ends.}
\label{fig:pipeline}
\end{figure}

Each activation chunk still passes through a chain of dependent stages, from HBM fetch and optical projection to chunk merge (\textbf{P1}--\textbf{P9} in Fig.~\ref{fig:local-attention}(b)). For OPT-175B, a chunk takes $F=1563.29$\,ns from fetch to merge, nearly eight times the 196.61\,ns activation-fetch interval at 8\,TB/s. Processing chunks serially would leave HBM idle most of the time, preventing the system from sustaining the HBM activation-fetch rate. OptiDecode therefore overlaps work at two levels, as illustrated in Fig.~\ref{fig:pipeline}.

\begin{figure*}[t]
  \centering
  \includegraphics[width=\textwidth]{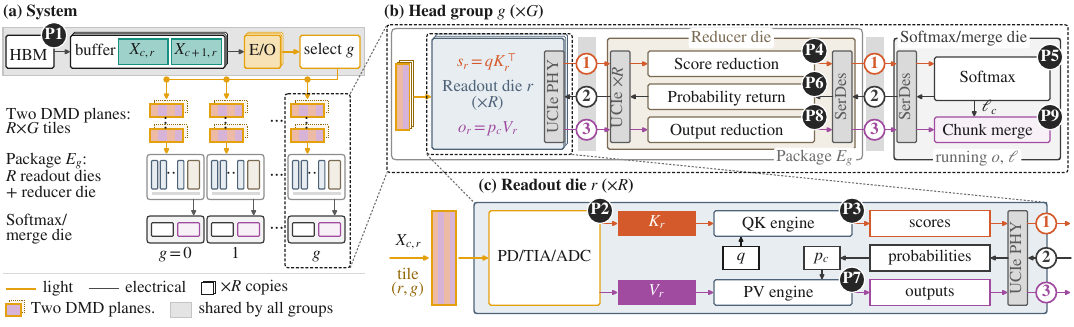}
  \caption{OptiDecode architecture. (a)~$R$ E/O converters and a group
  selector serve all $R\times G$ tiles. (b)~One head group: package $E_g$
  and its softmax/merge die. (c)~One readout die. P1--P9 and
  \textcircled{1}--\textcircled{3} follow Fig.~\ref{fig:local-attention}.}
  \label{fig:architecture}
\end{figure*}

\noindent\textbf{Across chunks.} The first level overlaps successive activation chunks. Each activation-buffer bank has two chunk slots: while the optics replay chunk $c$ from one slot, HBM fetches chunk $c+1$ into the other, and the digital engines and links continue processing earlier chunks (Fig.~\ref{fig:pipeline}). Let the chunk initiation interval $T$ be the time between the starts of successive chunks. For $N$ historical tokens split into $M=\lceil N/C\rceil$ chunks, the total processing time becomes $F+(M-1)T$ instead of $MF$ for serial execution; as the context $N$ grows, $F$ is amortized and throughput approaches one chunk every $T$. As shown in Fig.~\ref{fig:pipeline}, multiple chunks are in flight at once, with different resources working on different chunks.

\noindent\textbf{Across head groups.} The second level overlaps head groups within a chunk. Activation replay projects one head group at a time. However, a head group does not need to wait for the other groups to finish projection: 
as soon as a group's optical projection and readout (\textbf{P2}) finish, its readout dies can begin partial score computation (\textbf{P3}) while the optics project the next group (Fig.~\ref{fig:pipeline}). For OPT-175B at 8\,GBaud, each head group begins attention
as soon as its 16\,ns replay step completes, without waiting
for the full 192\,ns replay across all groups. 

\noindent\textbf{Matching the HBM activation-fetch rate.} The system starts a new chunk once HBM has delivered it and every hardware
resource has finished its work on the previous chunk. With $I_k$ the per-chunk service time of resource $k$ (including all stages and head groups that use it), the achievable chunk initiation interval is
\begin{equation}
T=\max\bigl(t_X,\max_k I_k\bigr).
\label{eq:chunk-interval}
\end{equation}
Our goal is $T=t_X$ because we want the system to accept one new chunk every activation-fetch interval. Equation~\eqref{eq:chunk-interval} shows that this requires $\max_k I_k\leq t_X$: every resource, including the optics, digital engines, reducer dies, and links, must finish its work for one chunk within one activation-fetch interval, leaving HBM as the only throughput bottleneck. This criterion bounds each resource's service time, not a chunk's end-to-end latency. Sustaining the overlap also requires buffering: each chunk stays in its activation-buffer bank until its last replay, and each readout die keeps its partial values until the attention probabilities return. Section~\ref{sec:architecture} organizes and provisions these resources.

\section{System Architecture}
\label{sec:architecture}

\subsection{Organization}
\label{sec:organization}
Fig.~\ref{fig:architecture} shows the OptiDecode architecture and maps the pipeline stages in Section~\ref{sec:dataflow} onto hardware. An $R\times G$ grid of optical tiles is fed by an activation input path shared across head groups, while each head group has its own readout and digital attention hardware.

\noindent\textbf{Shared activation and optical input path.} HBM fetches each activation chunk once (\textbf{P1}) into the on-chip activation buffer's $R$ double-slotted banks, one per input slice (Fig.~\ref{fig:architecture}(a)). Each bank drives one E/O converter, so $R$ converters serve all $G$ head groups via the group selector, as described in Section~\ref{sec:activation_replay}. Each tile uses two DMD planes: one serves the current layer while the other loads the next layer’s KV projection weights. Since reconfiguration latency is fixed while layer execution time grows with context length, this overlap hides reconfiguration at long contexts. 

\noindent\textbf{Readout and digital attention hardware.} Each optical tile has a corresponding readout die that digitizes its partial KV outputs through PDs, TIAs, and ADCs (\textbf{P2}) and stores partial keys $K_r$ and values $V_r$ in separate SRAM banks (Fig.~\ref{fig:architecture}(c)). Its QK engine computes partial scores from $K_r$ and $q$ (\textbf{P3}); its PV engine computes partial outputs from $V_r$ and the returned probabilities $p_c$ (\textbf{P7}); three message buffers hold the exchanged scores, probabilities, and outputs. For each head group $g$, the $R$ readout dies and one reducer die are integrated into package $E_g$ and connected through UCIe links; a separate softmax/merge die connects to the reducer die through a SerDes link (Fig.~\ref{fig:architecture}(b)). 

The dataflow follows the three message exchanges in Fig.~\ref{fig:local-attention}(b). First, each readout die sends its partial scores ($s_r$, \textcircled{1}) to the reducer die, which sums them into complete scores (\textbf{P4}). The softmax engine then computes the attention probabilities $p_c$ and the chunk LSE $\ell_c$ from these scores (\textbf{P5}). Second, the attention probabilities (\textcircled{2}) return to all $R$ readout dies (\textbf{P6}), where the PV engines compute partial outputs (\textbf{P7}). Third, the readout dies send their partial outputs ($o_r$, \textcircled{3}) to the reducer die, which sums them into the chunk output $o_c$ (\textbf{P8}). The chunk-merge engine combines $(o_c,\ell_c)$ with the running attention output and LSE $(o,\ell)$ using online softmax (\textbf{P9}). After all chunks are processed, the final attention result is returned to the host. At no point is partial KV gathered across dies. 

\noindent\textbf{Buffering for pipelined execution.} Since new chunks enter the pipeline before earlier chunks finish, intermediate data must remain buffered until their last use. Each activation-buffer bank has two 256\,KiB chunk slots and 16.4\,TB/s of read bandwidth to support replay. We size the other buffers from data lifetimes: data that live for $\tau$, with chunks starting every $T$, need at least $\lceil \tau/T\rceil$ chunk slots. For example, partial values are produced in \textbf{P2} but released only after \textbf{P7}; their 768.9\,ns lifetime requires seven 128\,KiB slots at the target chunk initiation interval of the provisioning workload. We allocate eight slots (1\,MiB) per readout die for partial values; with partial keys and message buffers, each readout die has 1.75\,MiB of SRAM.

\subsection{Provisioning}
\label{sec:provisioning}
We provision OptiDecode to sustain the HBM activation-fetch rate rather than to minimize the latency of an individual chunk. Since the target is $T=t_X$ with $t_X=Cdb/B_{\mathrm{HBM}}$ (Section~\ref{sec:pipeline}), a resource performing $w_k$ units of work per chunk needs a processing rate $\mu_k\ge w_k/t_X$. We use two provisioning rules to set the configuration.

\begin{table}[h]
  \centering
  \caption{OptiDecode hardware configuration.}
  \label{tab:provisioning}
  \footnotesize
  \setlength{\tabcolsep}{5pt}
  \renewcommand{\arraystretch}{1.05}
  \begin{tabular}{@{}l r l@{}}
    \toprule
    Component & Value & Detail \\
    \midrule
    \multicolumn{3}{@{}l}{\emph{Shared} ($\times1$)} \\
    HBM bandwidth            & 8\,TB/s       & B200 \\
    Chunk size $C$           & 128           & tokens \\
    \midrule
    \multicolumn{3}{@{}l}{\emph{Per input slice} ($\times R$)} \\
    Activation-buffer bank   & 512\,KiB      & 2 chunk slots \\
    \quad read bandwidth     & 16.4\,TB/s    & \\
    E/O converter            & 8\,GBaud      & 2{,}048 channels \\
    \quad wavelengths $N_\lambda$ & 1        & ${>}1$ when scaling \\
    \midrule
    \multicolumn{3}{@{}l}{\emph{Per optical tile and readout die} ($\times RG$)} \\
    Optical tile             & $2{,}048\times2{,}048$ & 2 DMD planes \\
    PD/TIA/ADC               & 8\,GS/s       & 2{,}048 channels \\
    QK engine                & 2.3\,TOPS     & INT8 \\
    PV engine                & 2.3\,TFLOP/s  & FP32 \\
    Readout-die SRAM         & 1.75\,MiB     & $K_r$, $V_r$, messages \\
    UCIe link                & 76.8\,GB/s    & to reducer die \\
    \midrule
    \multicolumn{3}{@{}l}{\emph{Per head group} ($\times G$)} \\
    Score reduction bank     & 9\,Gresults/s & INT32 \\
    Output reduction bank    & 9\,Gresults/s & FP32 \\
    Softmax engine           & 9\,Gscores/s  & \\
    Chunk-merge engine       & 28\,GOPS      & \\
    SerDes link              & 89.6\,GB/s    & to softmax/merge die \\
    Reducer-die SRAM         & 256\,KiB      & \\
    Softmax/merge-die SRAM   & 128\,KiB      & \\
    \midrule
    \textbf{Total SRAM}      & \textbf{133.5\,MiB} & \\
    \bottomrule
  \end{tabular}
\end{table}

\textit{First, processing and communication rates are sized for the shortest activation-fetch interval among the supported workloads.} We target MHA models with at least 30B parameters at an HBM bandwidth of 8\,TB/s, the B200 operating regime~\cite{nvidiaHGXB200Memory}. Since $t_X$ grows with the hidden dimension $d$, the supported model with the smallest hidden dimension ($d=7{,}168$) at 8\,TB/s sets the most demanding rate requirement. 

\textit{Second, hardware instance counts are sized for the largest projection mapping.} The supported models with the largest hidden dimension require $R=6$ input slices and $G=12$ head groups (OPT-175B). So, we provision six activation-buffer banks and E/O converters, a $6\times 12$ grid of optical tiles, and the corresponding readout dies, reducer dies, and softmax/merge dies. This configuration is fixed across models; models with smaller $d$ use only the required subset.

Table~\ref{tab:provisioning} summarizes the resulting per-instance configuration. Each $2{,}048\times2{,}048$ optical tile uses two DMD planes,
with 2{,}048 input channels modulated at 8\,GBaud and
2{,}048 output channels sampled at 8\,GS/s. The default configuration uses one wavelength; Section~\ref{sec:bw-scaling} adds wavelengths for HBM-bandwidth scaling. The $D=2{,}048$ tile dimension is a design target
informed by discussions with optical technology experts, rather than the specification of any particular commercial DMD. Each readout die provides 2.3\,TOPS of INT8 QK computation and 2.3\,TFLOP/s of FP32 PV computation. For each head group, the score and output reduction banks are provisioned at 9\,Gresults/s, the softmax engine at 9\,Gscores/s, and the chunk-merge engine at 28\,GOPS (counting three arithmetic operations per merged output element). Each readout--reducer connection uses a 64-lane-per-direction UCIe advanced-package module at 12\,GT/s, providing 96\,GB/s of raw bandwidth per direction~\cite{ucie_hc2023}. Each reducer--softmax connection uses eight 112\,Gb/s four-level pulse-amplitude modulation (PAM4) SerDes lanes per direction, providing 112\,GB/s of raw bandwidth~\cite{oif_xsrplus_2024,cadence_112g_xsr}. Assuming 80\% link utilization, their effective bandwidths are 76.8\,GB/s and 89.6\,GB/s per direction, respectively. With 512\,KiB per activation-buffer bank, 1.75\,MiB per readout
die, 256\,KiB per reducer die, and 128\,KiB per softmax/merge
die, the complete configuration requires 133.5\,MiB of on-chip SRAM.

\section{Evaluation}
\label{sec:evaluation}

\begin{figure*}[t]
  \centering
  \includegraphics[width=\textwidth]{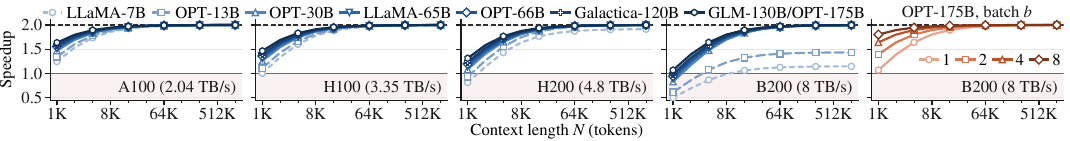}
  \caption{Decode-attention speedup over the ideal GPU KV-read baseline
across context lengths. Each comparison assigns OptiDecode
the same peak HBM bandwidth as the corresponding GPU.
The dashed line marks the $2\times$ speedup upper bound.}
  \label{fig:eval-hero}
\end{figure*}

\subsection{Methodology}
\label{sec:eval_method}
\noindent\textbf{Workloads.} We evaluate eight MHA models with head dimension $d_h=d_K=d_V=128$:
LLaMA-7B and 65B~\cite{touvron2023llama}, OPT-13B, 30B, 66B, and 175B~\cite{zhang2022opt},
Galactica-120B~\cite{taylor2022galactica}, and
GLM-130B~\cite{zeng2023glm130b}, from $d=4{,}096$ ($R=2$, $G=4$) to
$d=12{,}288$ ($R=6$, $G=12$). Models of 30B and above ($d\ge7{,}168$) are
the targets of OptiDecode's hardware provisioning (Table~\ref{tab:provisioning}); LLaMA-7B and OPT-13B test the limits of the provisioned hardware.
We sweep the context length $N$, the number of cached historical tokens excluding the current query token, from 1K to 1M tokens ($\mathrm{K}=2^{10}$, $\mathrm{M}=2^{20}$). We model one decode step with one query token, at a default batch size of 1.

\noindent\textbf{Baseline.} Our primary baseline is an \emph{ideal GPU KV read}: it streams
the $2Ndb$-byte INT8 historical KV cache at peak HBM bandwidth,
with all attention computation hidden, taking
$2Ndb/B_{\text{HBM}}$. The comparison covers only historical-cache
traffic and its attention computation, with equal peak HBM
bandwidth for both OptiDecode and the target GPU. This lower bound provides an ideal
GPU baseline for Figs.~\ref{fig:eval-hero},
\ref{fig:eval-stage-budget}, \ref{fig:bw-scaling},
\ref{fig:eval-sensitivity}, and
\ref{fig:eval-ablation-cumulative}.

We also measure GPU per-layer latency
(Fig.~\ref{fig:e2e-latency}) and energy
(Section~\ref{sec:eval-energy}) to quantify the gap from this ideal bound. We benchmark SGLang's FlashInfer decode-attention
path~\cite{zheng2024sglang,ye2025flashinfer} with FP8 KV caches and INT8 weights
on A100, H100, H200, and B200.
For projections and FFN, we select the faster INT8 implementation:
weight-only with \texttt{torch.compile}-fused dequantization
on A100--H200, and INT8$\times$INT8 cuBLASLt GEMMs on B200.
We flush L2 before each run, time per-layer attention and FFN
blocks with CUDA events, and report medians over
100 runs~\cite{nvidiaCudaEvents}. GPU energy is measured using
NVML's board energy counter~\cite{nvidiaNVMLEnergy} over
at least 3\,s of back-to-back runs.

\subsection{Decode Attention Speedup}
\label{sec:latency}
\textbf{Speedup across models and bandwidths.} Fig.~\ref{fig:eval-hero} shows that every target model (30B and above) reaches
$1.92$--$1.99\times$ at 32K tokens on all four bandwidths and
approximately $1.98\times$ or higher at 128K. This comparison isolates historical-cache traffic, the only HBM traffic OptiDecode changes, which grows with context length, while weight traffic remains fixed per layer. Once the pipeline is full, OptiDecode finishes a
chunk every activation-fetch interval ($t_X$), half the time the GPU takes
to read that chunk's KV ($2t_X$). 

The remaining gap to $2\times$ speedup is due to pipeline-fill overhead, which longer histories amortize. As $d$ increases, activation-fetch interval (\textbf{P1}) and optical replay (\textbf{P2}) grow, while digital-stage delays (\textbf{P3}--\textbf{P9}) remain unchanged because head groups use dedicated hardware in parallel (Section~\ref{sec:architecture}). As a result, activation-fetch interval grows faster than the fill overhead, allowing larger-$d$ models to approach $2\times$ at shorter histories.

\noindent \textbf{Smaller models.} Smaller-$d$ models gain less, especially on B200: at 32K tokens, LLaMA-7B and
OPT-13B reach $1.84$--$1.97\times$ on A100--H200 but $1.10\times$ and
$1.38\times$ on B200. Their narrower activations reduce the activation-fetch interval $t_X$, and B200's high HBM bandwidth reduces it further, below the minimum chunk initiation interval at which the digital engines (provisioned for larger models, Section~\ref{sec:provisioning}) can accept chunks at activation-fetch rate. 

\begin{figure}[h]
  \centering
  \includegraphics[width=\columnwidth]{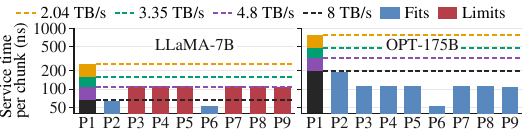}
  \caption{Per-chunk service
    time of each pipeline stage. P1 is the HBM activation fetch ($t_X$), shown for HBM bandwidths of different GPUs, and extended as budget lines.}
  \label{fig:eval-stage-budget}
\end{figure}

\noindent \textbf{Per-stage bottlenecks.} Fig.~\ref{fig:eval-stage-budget} shows each stage's per-chunk service time
($I_k$). With each engine and link operating at the fixed rate specified
in Section~\ref{sec:provisioning}, only HBM fetch (\textbf{P1}) and
optical replay (\textbf{P2}), shared across head groups, take
longer as $d$ grows. Larger-$d$ models use more of the provisioned
head groups in parallel, each with its own digital engines,
so the work per engine and service times of \textbf{P3}--\textbf{P9}
remain unchanged. Fig.~\ref{fig:eval-stage-budget} shows that the digital engines, provisioned
for models with at least 30B parameters ($d=7{,}168$), cannot keep pace with
LLaMA-7B's ($d=4{,}096$) activation-fetch rate at H200 and B200 bandwidths
(4.8 and 8\,TB/s). For OPT-175B, every stage keeps pace with
HBM at all bandwidths shown.

\noindent\textbf{Batching.} Sequences in a batch share the layer's resident
projection weights, so OptiDecode streams their activation chunks back to back through the same pipeline and pays the pipeline fill latency only once per batch,
while the GPU still reads a separate KV cache for each sequence. Batching therefore
helps most at short contexts, where the fill latency dominates
(Fig.~\ref{fig:eval-hero}, right). For OPT-175B on B200 at 1K tokens, speedup rises
from 1.07$\times$ at batch~1 to 1.80$\times$ at batch~8.

\begin{figure}[h]
  \centering
  \includegraphics[width=\columnwidth]{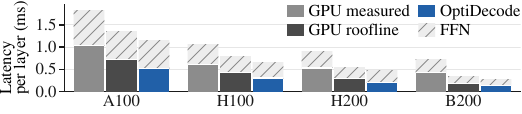}
  \caption{Per-layer decode latency (OPT-175B, $N{=}32$K, batch~1).
  Attention includes weight streaming from HBM.}
  \label{fig:e2e-latency}
\end{figure}

\noindent \textbf{Per-layer latency.} Fig.~\ref{fig:e2e-latency} breaks per-layer decode latency into attention
and FFN for OPT-175B at 32K tokens and batch~1. We compare OptiDecode
against two GPU references: measured execution on real hardware, and a
roofline that reads weights and KV at peak HBM bandwidth.
Unlike Fig.~\ref{fig:eval-hero}, OptiDecode here pays for weight
loading, and the FFN runs at the GPU roofline speed.
OptiDecode's attention is $1.99$--$3.30\times$ faster than measured GPU
attention, and the full layer is $1.18\times$ faster than
the GPU roofline.

\subsection{Scaling with HBM Bandwidth}
\label{sec:bw-scaling}

\begin{figure*}[t]
  \centering
  \includegraphics[width=\textwidth]{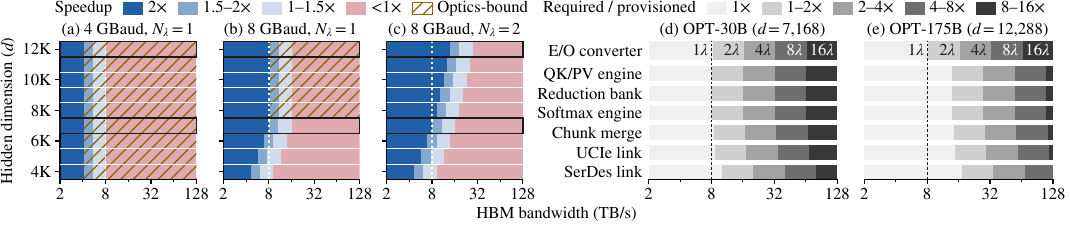}
    \caption{HBM-bandwidth scaling. (a)--(c)~Speedup over the ideal  KV-read baseline: $2\times$ regions are HBM-bound, hatched regions
optics-bound, and the rest digital-bound. (d)--(e)~Required
resource rates relative to provisioned rates.}
  \label{fig:bw-scaling}
\end{figure*}

\begin{figure*}[t]
  \centering
  \includegraphics[width=\textwidth]{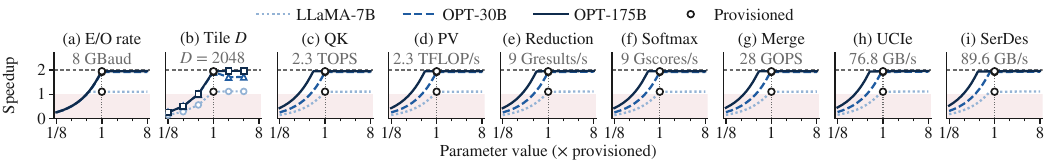}
  \caption{Parameter sensitivity of OptiDecode's speedup at 32K tokens and
8\,TB/s. Shading marks
speedup below $1\times$.}
  \label{fig:eval-sensitivity}
\end{figure*}

\textbf{Bottleneck regions.} Fig.~\ref{fig:bw-scaling}(a)--(c) shows how increasing HBM
bandwidth shifts the bottleneck. The optical and
digital \emph{bandwidth limits} are the highest HBM bandwidths
that activation replay and the digital engines and links,
respectively, can sustain. Below both limits, OptiDecode is \emph{HBM-bound}
and approaches $2\times$ speedup at long contexts (dark blue, Fig.~\ref{fig:bw-scaling}(a)--(c)).
Above the lower limit, the corresponding resource sets the chunk initiation
interval: hatched regions in Fig.~\ref{fig:bw-scaling}(a)--(c) are \emph{optics-bound}, while unhatched
regions below $2\times$ are \emph{digital-bound}.

At 4\,GBaud (Fig.~\ref{fig:bw-scaling}(a)), the HBM-bound region ends at 4.10\,TB/s, beyond which optical activation replay is the bottleneck. At 8\,GBaud (Fig.~\ref{fig:bw-scaling}(b))
the optical limit rises to 8.19\,TB/s, which models with $d\geq8{,}192$ can sustain; smaller-$d$ models hit their digital limits earlier. Adding a second wavelength ($N_\lambda = 2$, Fig.~\ref{fig:bw-scaling}(c)) doubles
optical throughput, raising the optical limit to 16.38\,TB/s, while digital and link rates remain unchanged. These resources
therefore limit throughput before optical replay does, eliminating optics-bound region. Compared to Fig.~\ref{fig:bw-scaling}(b), the $2\times$ region (HBM-bound) extends from 8.19 to 13.80\,TB/s for $d=12{,}288$, reaching
the digital limit, but remains unchanged for smaller-$d$ models
already limited by digital processing. This study shows what optical, digital, and link resources must scale for OptiDecode to preserve its speedup as HBM bandwidth increases.

\noindent \textbf{Resource scaling.} Figures~\ref{fig:bw-scaling}(d)--(e) show how much each resource's
rate must increase over its provisioned value (in Table~\ref{tab:provisioning}) to remain HBM-bound for OPT-30B and 175B (darker is larger scaling). Both models need twice as many wavelengths for each doubling
of HBM bandwidth. OPT-175B requires smaller increases in digital
engine and link rates, because its wider activation chunks (larger-$d$) take longer to fetch, while per-resource service times remain unchanged, as in Fig.~\ref{fig:eval-stage-budget}.

\subsection{Ablation}
\label{sec:ablation}
\begin{figure}[h]
  \centering
  \includegraphics[width=\columnwidth]{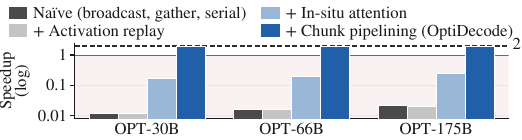}
  \caption{Ablation at 32K tokens and 8\,TB/s HBM
bandwidth.}
  \label{fig:eval-ablation-cumulative}
\end{figure}

Fig.~\ref{fig:eval-ablation-cumulative} adds each technique in turn, at 32K tokens and 8\,TB/s, relative to the ideal B200 KV-read baseline. The na\"ive design broadcasts activations to all head groups,
gathers and sums partial KV at each reducer, and processes
chunks serially; gathering and reducing KV consume over 95\% of its total latency, severely limiting its performance. \emph{Activation replay} reduces E/O converters by $G\times$
(72 to 6 for OPT-175B), with less than 1\% change in performance.
\emph{In-situ attention} removes the need to gather and reduce
partial KV, improving performance by $12$--$14\times$ over
the replay-only design, but chunks still execute serially.
\emph{Chunk pipelining} overlaps successive chunks, providing
a further $8$--$12\times$ improvement, reaching
$1.92$--$1.95\times$ the ideal GPU KV-read baseline performance.

\subsection{Parameter Sensitivity}
\label{sec:sensitivity}

Fig.~\ref{fig:eval-sensitivity} independently sweeps hardware parameters
from $1/8\times$ to $8\times$ their provisioned values
(Table~\ref{tab:provisioning}), at 32K tokens and 8\,TB/s. Increasing any single processing or link rate brings little benefit: OPT-30B and OPT-175B are already HBM-bound,
while LLaMA-7B has multiple digital bottlenecks, so accelerating
one resource leaves others limiting throughput. 

For OPT-30B, even small reductions in QK, PV, reduction,
or softmax rates make these resources bottlenecks. OPT-175B (larger-$d$)
tolerates larger reductions: individual digital and link rates
can fall to 47--58\% of provisioned values before limiting the
chunk initiation interval. Optical replay has no such advantage
for larger models: both become replay-limited below $0.98\times$
the provisioned modulation rate, whose requirement is independent
of $d$ (Eq.~\eqref{eq:replay-freq}).

Tile dimension $D$ impacts both replay time and hardware utilization. Halving $D$ doubles the number of head groups and approximately
halves speedup (Fig.~\ref{fig:eval-sensitivity}(b)). Increasing $D$ reduces replay steps
and can improve speedup in the optics-bound regions of
Fig.~\ref{fig:bw-scaling}(b), until HBM or digital processing
becomes limiting. At the evaluated operating point (8\,TB/s, $D=2048$, 8\,GBaud), however,
replay already keeps pace with HBM, so larger tiles offer little
benefit at fixed total digital and link rates. At $D=4096$, each tile accommodates 16 heads, so OPT-30B's
56 heads occupy four groups with only eight heads in the last.
This underutilization reduces speedup from $1.92\times$ to
$1.70\times$. Balancing the number of heads across groups provisioned with equal
hardware 
avoids this underutilization.

\subsection{Energy}
\label{sec:eval-energy}
We estimate energy with a parameterized first-order model that combines published component metrics~\cite{moon2023advanced,9499913,yokoyama202029,9997502,atef20148,swindlehurst20218,walden1999analog,sharma2025uciememory,rambus2019xsr,ti2018dlp650lnir,Spall:20,zhang2026skylight}. We report energy per historical token per attention layer
for one OPT-175B decode step at batch~1 and 32K context.
We use two GPU baselines: \emph{TDP-based}, which multiplies
vendor-specified TDP by the ideal KV-read time
at peak HBM bandwidth, excluding kernel overheads; and
\emph{measured}, which uses the board energy
counter~\cite{nvidiaNVMLEnergy} to measure SGLang's FlashInfer
decode-attention kernel~\cite{zheng2024sglang,ye2025flashinfer}.
Each comparison gives OptiDecode the same HBM bandwidth
as the GPU it is compared against.

\noindent \textbf{Against the TDP-based baseline.} At each GPU's peak HBM
bandwidth of 8, 4.8, and 2.04\,TB/s, OptiDecode's estimated energy is 3.31,
3.37, and 3.66\,\textmu J per historical token per layer, respectively. This
is 23\% below the 4.30\,\textmu J estimate for a Blackwell Ultra GPU
(8\,TB/s, 1,400\,W~\cite{aubrey2025}), 6\% below the
3.58\,\textmu J for an H200 SXM (4.8\,TB/s, 700\,W~\cite{nvidiaH200}), and
24\% below the 4.82\,\textmu J for an A100 SXM (2.04\,TB/s,
400\,W~\cite{nvidiaA100}).

\noindent \textbf{Against the measured baseline.} A100 PCIe, H200 SXM, and
B200 SXM consume 5.84, 5.76, and 5.85\,\textmu J per historical token
per layer, respectively, or $1.5$--$1.9\times$ their TDP-based
estimates. At
each GPU's HBM bandwidth (1.94, 4.8, and
8\,TB/s~\cite{nvidiaA100,nvidiaH200,nvidiaHGXB200Memory}), OptiDecode's
estimates of 3.68, 3.37, and 3.31\,\textmu J are 37--43\% lower than these GPU energy measurements.

\begin{figure}[h]
  \centering
  \includegraphics[width=\columnwidth]{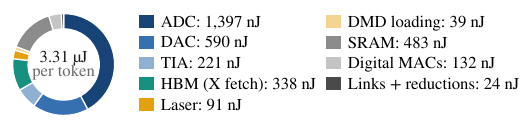}
  \caption{Modeled OptiDecode attention energy per historical token per layer (OPT-175B,
    32K tokens, 8\,TB/s HBM).}
  \label{fig:eval-energy-breakdown}
\end{figure}
 
\noindent \textbf{Energy breakdown.} Fig.~\ref{fig:eval-energy-breakdown} shows that most
of the 3.31\,\textmu J is spent converting signals between the electrical and
optical domains, not on the optics. The ADCs, DACs, and receiver amplifiers use
67\% (the ADCs alone 42\%), on-chip SRAM 15\%, and reading cached activations
from HBM only 10\%. The laser uses just 2.7\% and DMD loading 1.2\%.    Converters are therefore the largest opportunity for further savings within this model.

\section{Related Work}
\label{sec:related-work}

Prior systems reduce KV storage or traffic by regenerating part of the history's KV on demand. CAPTURE~\cite{lee2025throughput} and KVPR~\cite{jiang2025kvpr} keep part of the history as layer-input activations in host memory and regenerate its KV on the GPU while weights or the remaining KV cross PCIe. ALISA~\cite{zhao2024alisa} attends to a sparse subset of tokens and recomputes old offloaded KV once fetching it becomes slower. Kelle~\cite{xia2025kelle} stores layer-input activations instead of KV for tokens that most heads retain, and XQuant~\cite{tomar2025xquant} hides regeneration from quantized inputs behind weight loading, which on an H100 suffices only up to approximately 2.3K tokens (Llama-2-7B, 2-bit). Because digital KV regeneration has an arithmetic intensity of approximately $4d$ operations per byte of activation, far above a GPU's compute-to-bandwidth ratio, these systems regenerate only what a slower transfer can hide; otherwise they affect decode performance. OptiDecode bypasses this limit by moving KV regeneration to FSO computing, regenerating the entire KV history at the HBM activation-fetch rate.

Most photonic accelerators for deep learning use integrated silicon
photonics, both in research
designs~\cite{liu2019holylight,demirkiran2023electro,liu2025lightml} and in
commercial chips~\cite{ahmed2025universal,lightmatter_envise,lightelligence_pace2}.
Prior works extend silicon
photonics to transformers as well~\cite{zhu2024lightening,karimi2026scarlet,chang2025p}. Silicon photonic designs, however, are hard to scale to LLM-sized matrices: an
\(n_{\mathrm{mesh}}\times n_{\mathrm{mesh}}\) mesh cascades $O(n_{\mathrm{mesh}})$ devices along each optical path and tunes $O(n_{\mathrm{mesh}}^2)$
waveguide elements, so optical loss, crosstalk and phase-tuning power grow with
matrix size~\cite{al2022scaling,shafiee2024analysis}. Free-space optics
overcomes this limitation. Lumai's Iris Nova~\cite{lumai2026iris} applies free-space optics to
compute-bound prefill, a natural fit for FSO's extremely high compute throughput.

\section{Future Directions}
\label{sec:discussion}
OptiDecode reduces dense MHA's historical-cache traffic
while retaining independent KV heads and attention over
the full history. Head sharing, sparsity, and latent representations offer
other algorithmic approaches to efficient attention, opening
opportunities to extend OptiDecode. Although GQA~\cite{ainslie2023gqa} shares KV across
heads, reducing the opportunity for activation caching, sparse MHA offers a complementary extension: OptiDecode
could regenerate KV from cached activations for tokens
selected by many heads, while reading KV directly for
tokens selected by few heads. Co-designing token selection
and such a hybrid caching paradigm is a promising future direction. A further extension could combine sparse retrieval with
latent caching~\cite{liu2024deepseek,chang2025palu}, using optical regeneration to reduce the
digital cost of reconstructing KV from selected tokens'
compact latent vectors where reconstruction is required.

\section{Conclusion}
OptiDecode revisits what long-context MHA decoding should cache.
Caching activations instead of KV halves historical-cache storage
and traffic, provided KV regeneration keeps pace with HBM.
Free-space optics supplies this computation using resident
projection weights; activation replay avoids broadcast,
in-situ attention prevents partial KV communication, and chunk pipelining
sustains the HBM fetch rate. OptiDecode shows
$1.92$--$1.99\times$ decode-attention speedup over the ideal GPU
KV-read baseline at equal peak HBM bandwidth, across evaluated
30B--175B MHA models at 32K-token contexts. This shows how emerging technologies can make otherwise
prohibitive design choices efficient, provided the dataflow
and architecture effectively exploit their advantages.

\begin{acks}
We used Anthropic's Claude and OpenAI's ChatGPT
to assist with manuscript drafting and editing, code development,
figure preparation, and verifying calculations.
\end{acks}

\bibliographystyle{ACM-Reference-Format}
\bibliography{refs}

\end{document}